\documentclass[review]{elsarticle}

\usepackage{lineno,hyperref}
\modulolinenumbers[5]

\journal{Journal of \LaTeX\ Templates}
\usepackage[T1]{fontenc}
\usepackage{ae,aecompl}
\usepackage{color}

\usepackage{graphicx}	
\usepackage{amsmath}	
\usepackage{amssymb}	

\begin{document}

\begin{frontmatter}

\title{Direct reconstruction of dynamical dark energy from observational Hubble parameter data\tnoteref{mytitlenote}}
\tnotetext[mytitlenote]{Fully documented templates are available in the elsarticle package on \href{http://www.ctan.org/tex-archive/macros/latex/contrib/elsarticle}{CTAN}.}

\author{Zhi-E Liu}
\address{College of Physics and Electronic Engineering, Qilu Normal University, Jinan, 250014, China}

\author[foryu]{Hao-Ran Yu}
\author[mymainaddress]{Tong-Jie Zhang\corref{mycorrespondingauthor}}
\cortext[mycorrespondingauthor]{Corresponding author}
\ead{tjzhang@bnu.edu.cn}

\author[mysecondaryaddress]{Yan-Ke Tang}
\address[foryu]{Kavli Institute for Astronomy and Astrophysics, Peking University, Beijing 100871, China}
\address[mymainaddress]{Department of Astronomy, Beijing Normal University, Beijing, 100875, China}
\address[mysecondaryaddress]{Department of Physics, Dezhou University, Dezhou, 253023, China}


\begin{abstract}
Reconstructing the evolution history of the dark energy
equation of state parameter $w(z)$ directly from observational
data is highly valuable in cosmology, since it contains substantial
clues in understanding the nature of the accelerated expansion of
the Universe. Many works have focused on reconstructing
$w(z)$ using Type Ia supernova data, however, only a few studies pay
attention to Hubble parameter data. In the present work, we explore
the merit of Hubble parameter data and make an attempt to
reconstruct $w(z)$ from them through the principle component
analysis approach. We find that current Hubble parameter data
perform well in reconstructing $w(z)$; though, when compared to
supernova data, the data are scant and their quality is worse.
Both $\Lambda$CDM and evolving $w(z)$
models can be constrained within $10\%$ at redshifts
 $z \lesssim 1.5$ and even $5\%$ at redshifts 0.1 $\lesssim$ z $\lesssim$ 1 by using simulated $H(z)$
data of observational quality.
\end{abstract}

\begin{keyword}
\texttt{Dark Energy }\sep Reconstruction\sep Nonparametric Model \sep Observational Hubble Parameter Data
\end{keyword}

\end{frontmatter}


\section{Introduction}
\label{sec:intro}

There have been a handful of works reconstructing the dark energy equation of state parameter
$w(z)$, defined as $w(z)=P/\rho$, where $P$ and $\rho$
represent the pressure and energy density, from observed distance
modulus data (e.g., Type Ia supernova (SN Ia) data). Basically, these works
can be divided into two categories: (i) Assume a particular model
about $w(z)$ so that the model quantities (e.g., luminosity
distance $d_{L}$) can be compared directly with the observations.
(ii) Characterize the data with an underlying model and
then reconstruct $w(z)$ numerically. For category (i), the
assumed model can be expressed as a parameterized form of a basis
representation or a function distribution representation. The
parametric forms are currently the most commonly used, which either
assume $w(z)$ is a constant or allow for a redshift
dependence, such as $w(z) = w_{0} - w_{1}z/(1 + z)$,
where $w_{0}$ and $w_{1}$ are constants
\cite{Chevallier200110,Linder200390}. Hojjati et al. \cite{Hojjati201004} used wavelets to study the
redshift evolution of $w(z)$. In that work the Principal
Component Analysis (PCA) technique, which has the advantage of detecting
local features such as bumps in $w(z)$, was applied to
decorrelate the wavelet coefficients. Huterer et al. \cite{Huterer200571} presented a
piecewise constant description of $w(z)$ by binning redshifts,
and estimated the corresponding band power. The PCA was also used to
make the band power uncorrelated so that the eigenmodes can be
written as linear combinations of redshift bins. Holsclaw et al. \cite{Holsclaw2010105}
introduced a non-parametric reconstruction method by modeling
$w(z)$ through a Gaussian process (GP) and estimating the GP
hyperparameters with a Bayesian inference and Markov chain Monte Carlo
sampling. Zhao et al. \cite{Zhao2012109} combined GP and redshift binning method to
feature $w(z)$, in which the $\omega$ value in each bin is
treated as free parameter to be determined. To make $w(z)$
differentiable for the calculation of the dark energy perturbations,
they interpolated between the bins with narrow tanh functions.For
category (ii), Daly et al. \cite{Daly2003597} developed a simple numerical method
to determine $w(z)$ from the data, which made no assumptions
about the underlying cosmological models. Their approach is
model-independent. However, the final results are noisier and highly
sensitive to the amount and quality of the available data. To
suppress the noise, Shafieloo et al. \cite{Shafieloo2006366} smoothed the SN Ia data by
applying a Gaussian kernel over the whole redshift range before
performing reconstruction. Majumdar et al. \cite{Majumdar2012375} expressed the distance
modulus in a parametric form, $\mu = 5
\log{[z(1+az)/(1+bz)]}+\mu_{0}$, and used the standard $\chi^2$
minimization to determine the parameters (\textit{a}, \textit{b},
$\mu_0$). Clarkson et al. \cite{Clarkson2010104} fitted the data before reconstruction using
eigenmodes which were transformed from a set of primal basis
functions with PCA.

With observational data getting richer and more diverse, recently researchers
have begun to study dark energy reconstruction with multiple sets of measurements.
Holsclaw et al. \cite{Holsclaw201184} extended their Gaussian process method \cite{Holsclaw2010105} to include
a diverse set of measurements: baryon acoustic oscillations (BAO), cosmic microwave
background measurements (CMB), and supernova data (SNe), while observational Hubble
parameter $H(z)$ data (OHD) were used by other researchers \cite{Majumdar2012375, Zhao2012109, Yu201388}.

So far, most research has focused on observational data acquired
from SNe, CMB, and BAO, especially from SNe, or their hypothetical
counterparts. On the contrary, the problem of reconstructing dark
energy from Hubble parameter data has attracted less attention of
the cosmology community. How valuable are OHD in modeling the
Universe? In fact, some pioneers have made explorative works in this
area. Ma et al. \cite{Ma2011730} studied the viability of constraining the
cosmological parameters using OHD and concluded that
as many as 64 data points are
sufficient to get a result comparative to SNe data.
Pan et al. \cite{Pan20101012} discovered that to some extent the OHD were better
poised to give information about the nature of dark energy than the
SN Ia data. Encouraged by these pioneering works, in this paper we
attempt to reconstruct dark energy solely with the help of OHD.

The paper is structured as follows. In section \ref{merit}, we analyze the advantage of OHD.
The reconstruction method we adopt is outlined in section \ref{methodology}.
Our results are presented in section \ref{result}, and finally we make some conclusions
in section \ref{conclusion}.

\section{Merit of the Hubble parameter}
\label{merit}

\begin{figure}[htbp]
\centering
\includegraphics[width=0.80\textwidth]{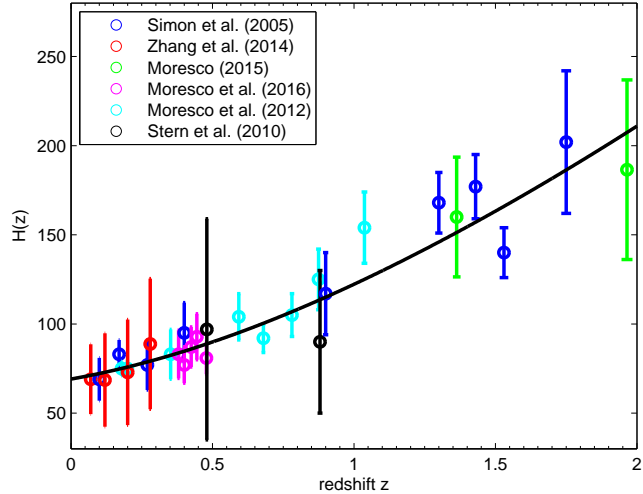}
\caption{ OHD data and the best-fit $\Lambda$CDM model. The black solid curve represents the $\Lambda$CDM model best-fitting the data, which gives $\Omega_{m} = 0.36$, $\Omega_{k} = -0.13$ and
$H_{0} = 69.1$\,km/s/Mpc.}
\label{Fig:od}
\end{figure}

Although, as previously mentioned, some authors have considered the
role of OHD in reconstructing dark energy with combined data sets,
few researchers have tried to use OHD solely to undertake this task \cite{Moresco2016}.
This is partially due to the smaller number and lower quality of OHD
compared to the distance module data. In contrast to the 580 SN Ia
data released by the supernova cosmology project \cite{Suzuki2012746}, to
date, the most comprehensive sample of OHD contains only 37 measurements (see Figs.~\ref{Fig:od}). They are collected from
\cite{Stern201002, Moresco201208, Busca2013552, Zhang201414, Blake2012405, Chuang2013435, Delubac2015574, Moresco2015, Moresco2016, Planck2014571}, most of which are compiled in \cite{Farooq2013766}. Current
OHD are obtained primarily by the cosmic chronometers method (30 measurements)
\cite{Stern201002, Moresco201208, Zhang201414, Moresco2015, Moresco2016}. Radial BAO peaks detection is another
method to extract $H(z)$ \cite{Blake2012405, Chuang2013435,Delubac2015574}. The cosmic chronometers method is affected by the
systematic biases in the age determination of galaxies, while for the radial BAO
method, the systematics stem from various distortion effects
subjected to the autocorrelation function. Since a fiducial flat
Lambda cold dark matter ($\Lambda$CDM) model and its parameters are
used to convert redshifts into distances and to gauge the comoving
BAO scales in the selected redshift slice \cite{Gaztanaga2009399}, the radial
BAO method is essentially model-dependent, and the possible circular
logic issue also contributes to the systematic errors in this method
\cite{Heavens2014113}. Since OHD measurements coming from different techniques may suffer from different systematic effects, the 30 measurements that come from cosmic chronometers are adopted in our analysis (see Fig.\ref{Fig:od}). They all have been compiled in Ref. \cite{Moresco2016}.

OHD (with mean relative error of $19.8\%$) have
generally larger error bars than SN Ia data (with
mean relative error of merely $0.55\%$). However,
OHD $H(z)$ have more direct and efficient power to constrain
$w(z)$ compared to SN Ia $d_L(z)$ data. $w(z)$
is written in terms of $d_L(z)$ as Equation (\ref{wzD}) \cite{Clarkson2010104}:
\begin{eqnarray}
\label{wzD}
\omega(z)&=&\{2(1+z)(1+\Omega_{k}D^2)D''-[(1+z)^2\Omega_{k}D'^2 \nonumber\\
& &+2(1+z)\Omega_{k}DD'-3(1+\Omega_{k}D^2)]D'\}/\{3\{(1+z)^2 \nonumber\\
& &\times[\Omega_{k}+(1+z)\Omega_{m}]D'^2-(1+\Omega_{k}D^2)\}D'\},
\end{eqnarray}
Here $\Omega_{m}$ and
$\Omega_{k}$ are normalized density parameters of matter and curvature and prime denotes the derivative with
respect to $z$. $D(z) = d_L(z)H_0/(c(1+z))$ is the normalized
comoving distance and $H_{0} = H(0)$ is the Hubble constant. Note that the second derivative of $D(z)$ is a prerequisite in Eq.~(\ref{wzD})
even in the case of $\Omega_{k} = 0$. We can decompose Eq.~(\ref{wzD}) into two simpler equations by introducing
the Hubble parameter $H(z)$ as the intermediate quantity, which can be expressed as Equation (\ref{DtoH}),
\begin{equation}
\label{DtoH}
H(z) = \frac{H_{0}}{D'(z)}\sqrt{1+[D(z)]^2\Omega_{k}}
\end{equation}
and Equation (\ref{wzH}),
\begin{equation}
\label{wzH}
w(z) = \frac{1}{3}\frac{2HH'(1+z)-3H^2+H_{0}^2\Omega_{k}(1+z)^2}{H^2-H_{0}^2\Omega_{m}(1+z)^3-H_{0}^2\Omega_{k}(1+z)^2}.
\end{equation}
Eq.~(\ref{wzH}) shows that to calculate $w(z)$ from $H(z)$, only the first derivative of $H(z)$ is needed.
The derivative operator functions as a high-pass filter and can thus heavily amplify the observational errors.
Therefore, to reconstruct $w(z)$, $H(z)$ is preferable to $d_{L}(z)$ when regarding the suppression of
the propagation of noise \cite{Maor200265}. Such an advantage of $H(z)$ makes it a promising method to reconstruct
$w(z)$ directly from $H(z)$ instead of $d_{L}(z)$, although the current quality of $H(z)$ data is comparatively poor.
It should be noted that Pan et al. \cite{Pan20101012} have given an estimate about the advantage of the first derivative of data over
the second derivative in reconstructing dark energy. They studied the efficacy of four different parameters:
the deceleration parameter, the equation of state of dark energy, the dark energy density, and the geometrical
parameters $Om(z)$, which can be expressed as Equation (\ref{omz}) \cite{Sahni200878}:
\begin{equation}
\label{omz}
Om(z) = \frac{\tilde{h}^{2}(z)-1}{(1+z)^{3}-1}, \tilde{h} = H(z)/H_{0}.
\end{equation}
In discriminating theoretical models of dark energy using supernova
data, they found that the geometrical parameters $Om(z)$, a
cosmological parameter constructed from the first derivative of the
data, for which the theoretical models of dark energy are
sufficiently distant from each other, performs best in
reconstructing dark energy from SNe data.

\section{Methodology}
\label{methodology}

Most existing reconstruction approaches making use of the observed distance modulus data can also be applied to OHD.
Here we apply the framework of weighted least square (WLS) and PCA \cite{Clarkson2010104}. Given a set of $H(z)$
data $X=(x_1,x_2,...,x_K)^T$, we fit them with a smooth analytical function, and then use it and its first derivative
to reconstruct $w(z)$ according to Eq.~(\ref{wzH}). Assume the covariance matrix $C$ of the data $X$
satisfies $C_{ii}=\sigma^2_i$ for $i=1,2,...,K$ and $C_{ij}=0$ for $i\neq j$. We start with a set of \textit{N} primary
basis functions $p_{n}(z)$, and their summation $\sum^N_{n=1}{a_{n}p_{n}(z)}$ is used to fit $X$ through the weighted least
square technique, with the weight matrix being $C^{-1}$. Minimizing the weighted squared
residual $\mathcal{R}=(X-PA)^TC^{-1}(X-PA)$ by solving $\partial \mathcal{R}/\partial A = 0$, we get the coefficient vector $A=(a_1,a_2,...,a_N)^T$ as shown in Equation (\ref{coeffa}):
\begin{equation}
\label{coeffa}
A = (P^TC^{-1}P)^{-1}P^TC^{-1}X,
\end{equation}
where the $K\times N$ measurement matrix $P=(p_1,p_2,...,p_N)$ is composed of the basis function set. Meanwhile, according to the least squares theory, the inverse covariance matrix of coefficient vector $A$ is given by Equation (\ref{invcova}):
\begin{equation}\label{invcova}
  Cov^{-1}(A)=P^TC^{-1}P.
\end{equation}

Then the PCA is implemented by calculating eigenvalues and eigenvectors of $Cov^{-1}(A)$: $Cov^{-1}(A)=E\Lambda E^T$, where $E=(e_1,e_2,...,e_N)$ is the matrix of eigenvectors and $\Lambda$ the diagonal matrix with diagonal entities
representing eigenvalues with decreasing order. A set of $N$ optimal basis functions $Q=(q_1,q_2,...,q_N)$ are produced by
transforming the primary functions through $E$: $Q=PE$. That is given by Equation (\ref{transform}),
\begin{equation}
\label{transform}
q_n(z) = \sum^N_{m=1}e_{mn}p_m(z), n=1...N,
\end{equation}
where $e_{mn}$ is the \emph{m}th component of the \emph{n}th eigenvector.
The first $M$ ($M<N$) eigenfuncitons $Q'=(q_1,q_2,...,q_M)$ are
usually of interest since they are the principal components and represent the main features of the data.
We use them to fit \emph{X} once more with WLS and get new coefficients $A'=(a'_1,a'_2,...,a'_M)^T$ given by Equation (\ref{coeffap}):
\begin{equation}
\label{coeffap}
A' = (Q'^TC^{-1}Q')^{-1}Q'^TC^{-1}X.
\end{equation}
Then the function $\sum^M_{n=1}{a'_{n}q_{n}(z)}$ gives the analytic
representative of the data \emph{X}. Thus, each choice $\{N,M\}$
will give a particular reconstruction of $w(z)$.

To balance the risk of getting a wrong $w(z)$ and the risk of
over-fitting, we adopted the formalism discussed in \cite{Clarkson2010104}. That is, a combined information criterion is used to cull out the most suitable combinations of $\{N,M\}$ with $N \in [2,10]$ and $M \in [2,N]$, which is defined as Equation (\ref{cic}):
\begin{equation}
\label{cic}
\rm{CIC} = (1-\emph{s})AIC+\emph{s}BIC,
\end{equation}
where AIC is the Akaike information criteria and BIC the Bayesian criteria.
This framework does not single out one particular combination of $\{N,M\}$ that possesses minimum CIC, but selects a family of $\{N,M\}$ reconstructions which satisfy Equation (\ref{cicsup}):
\begin{equation}
\label{cicsup}
\rm{CIC} < \rm{CIC}_{\rm{min}}+\kappa,
\end{equation}
where $\kappa$ is a constant being set to 5 through the paper. From the selected family of $\{N,M\}$ reconstructions the averaged $w(z)$ and $1-\sigma$ errors are inferred. Thus, $\kappa$ controls the size of the family of $\{N,M\}$ combinations and $s$
gives the weight for the importance of the Bayesian criteria relative to the Akaike criteria.

Both OHD (denote by $H_o(z)$ hereafter) and simulated $H(z)$ data (denote $H_s(z)$ hereafter) are tested in our experiments. Moresco\cite{Moresco2015} performed a $H(z)$ simulation of a Euclid-like survey to forecast the expected improvement on estimating cosmological parameters with future data. Here, we get $H_s(z)$ based on currently observed data by using the method
proposed in \cite{Yu201388}. We assume $\Omega_{m} = 0.30$ and $\Omega_{k} = 0$ and the Hubble constant $H_{0}$ is set to the newest value of $67.4$\,km/s/Mpc \cite{Planck2014571}. In this method, we take the relative error $\sigma_H/H$
as a random variable which satisfies the \textit{z}-independent Nakagami distribution $f_m(x;m,\Omega)$. By fitting OHD, we get the
following distribution parameters: $m=0.63$ and $\Omega = 0.038$. The final $H_s(z)$ data
should follow the Gaussian distribution $H_{s}(z) \sim N(H_{\bigstar}(z), \sigma_H)$, where $H_{\bigstar}(z)$ is the fiducial \textit{H}
which gives the simulated data. In total, 400 data points with $z$ evenly distributed in the range of [0, 2.0] are generated.
The uncertainties of the $H_s(z)$ increase with \textit{z}. As a result, the simulated data set is
qualitatively comparable with observations.

 \section{Results}
 \label{result}

 \subsection{Results on simulated data}
 \label{resultsim}

\begin{figure}
\centering
\includegraphics[width=9cm]{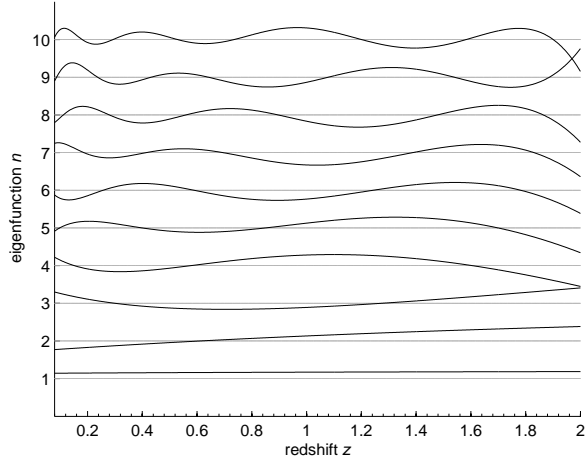}
\caption{The first 10 eigenfunctions for the $H_s(z)$ data, with \textit{N} = 10. Note the numbers aligning the ordinate mark the order of these eigenfunctions other than their own values.
\label{Fig:eigf}}
\end{figure}

\begin{figure}
\centering
\includegraphics[width=0.80\textwidth]{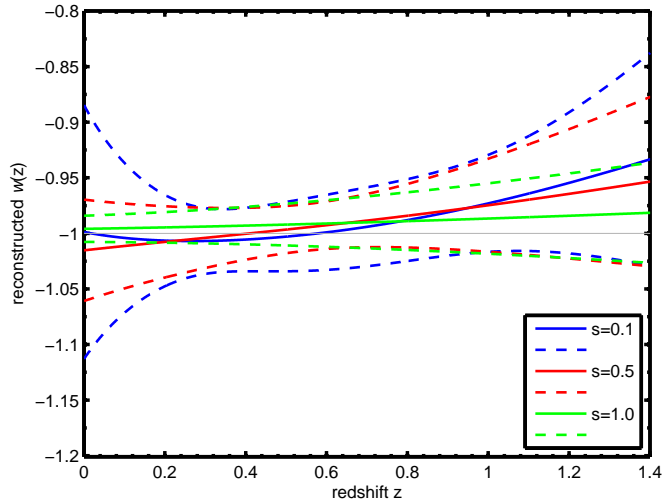}
\caption{Reconstruction of $w(z)$ using $H_s(z)$ data. The solid curves
represent the reconstructed $w(z)$ and the dashed curves represent 1-$\sigma$ errors. The gray horizontal line denotes $w=-1$}
\label{Fig:shd2w}
\end{figure}

We first test the power of the $H_s(z)$ data with the above technique.
The primary basis functions are $p_{n}(z) = [z/(1+z)]^{n-1}$, $n = 1\sim N$. Other types
of functions, such as  $z^{n-1}$ and $1/(1+z)^{n-1}$, are also tested but led to similar results. The function
set $z^{n-1}$ leads to slightly larger reconstruction errors, while the reconstruction errors given by $[1/(1+z)]^{n-1}$
are slightly smaller than $[z/(1+z)]^{n-1}$. In Fig.\ref{Fig:eigf}, we show the first 10 normalized eigenfunctions obtained by fitting the $H_s(z)$ data with PCA, with the underlying model being \textit{w} = -1 and \textit{N} = 10. We find that, for a fixed \textit{N}, the shapes of eigenfunctions vary from smooth to oscillatory. The fiducial model used in this procedure has the same parameters as
those used in the simulation procedures, i.e., $\{\Omega_{m},\Omega_{k}, H_{0}\} = \{0.30, 0.0, 67.4\}$
and $w_{\rm{fiducial}} = -1$. We set $\kappa = 5$ and study three cases $s$ = 0.1, 0.5, and 1.0,
respectively. Fig.\ref{Fig:shd2w} shows
the reconstructed $w(z)$ (denote by $w_{Hs}(z)$ hereafter). On the whole, the results are quite good for all $s$ values.
Fig.\ref{Fig:shd2w} shows that $s$ takes us
from conservative models where $s$ = 1 to more wild models where $s$ = 0.1. This is due to the fact that
models minimizing the BIC criteria tend to become smooth with tight error bars,
while models with low AIC values are oscillatory and have large errors.

\begin{figure}
\centering
\includegraphics[width=0.8\textwidth]{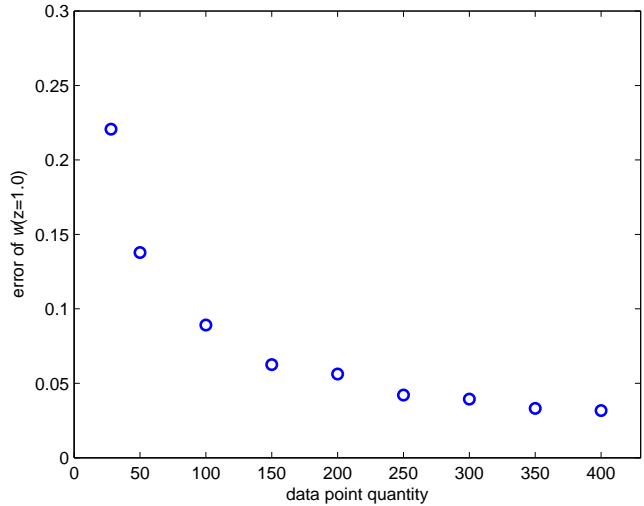}
\caption{The reconstruction errors of $w(z)$ at $z=1$.
Here $w(z)$ are reconstructed by $H_s(z)$ data with different number of data points.
We set $s=1.0$ and $\kappa = 5$.}
\label{Fig:werrz1}
\end{figure}

It is expected that the number of data points should have a crucial
impact on the quality of the reconstruction. As shown in
Fig.\ref{Fig:werrz1}, the reconstruction error drops drastically
with an increasing number of the simulated data points, except for the
number greater than 200 or so. For
200 or more data points, the resulting errors are comparable to what
are shown in Fig.\ref{Fig:shd2w}. We can see that the currently
observed 30 $H_o(z)$ data points, with current measure quality, are
really too sparse. We expect more OHD data points to be measured in future. On the other hand, 400 data points
seem to be enough for improving the reconstruction quality, and further
increasing the number of data points will merely lead to a marginal
improvement. Further improvement on the reconstruction result needs
further improvement on data quality.

\begin{figure}
\centering
\includegraphics[width=0.8\textwidth]{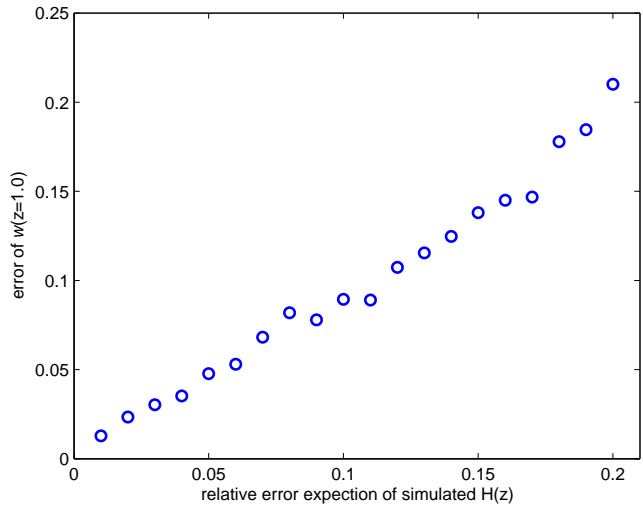}
\caption{The reconstruction errors of $w(z)$ at $z=1$.
Here $w(z)$ are reconstructed by $H_s(z)$ data with different quality but fixed 30 data points.
We set $s=1.0$ and $\kappa = 5$.}
\label{Fig:werrherr}
\end{figure}

However, it is unlikely that future surveys will provide hundreds of $H(z)$ measurements. On the contrary, current measurements are basically systematic-dominated, so there is space for improvement in the direction of getting smaller errorbars, even at higher-z. Thus, we give the $H(z)$ simulation by keeping the number 30 of data points fixed but reducing $H_s(z)$ errors. How the reconstructed $w(z)$ errors vary with different $H_s(z)$ errors is shown in Fig.\ref{Fig:werrherr}. We can see that, to make the reconstructed $w(z)$ error less than 0.05 at $z=1$, which is equivalent to the reconstruction accuracy obtained by 400 $H_s(z)$ points of current measurement quality (see Fig.\ref{Fig:werrz1}), the relative error of $H_s(z)$ should achieve $4\%$. Obviously this is a big challenge for future surveys. We alternatively test 50 $H_s(z)$ points and find the needed relative $H_s(z)$ error should be $6\%$, and for 100 $H_s(z)$ points the relative error $12\%$ is necessary. Considering the feasibility of future implement, 100 OHD measurements with quality of $12\%$ are desirable.

\begin{figure}
\centering
\includegraphics[width=9cm]{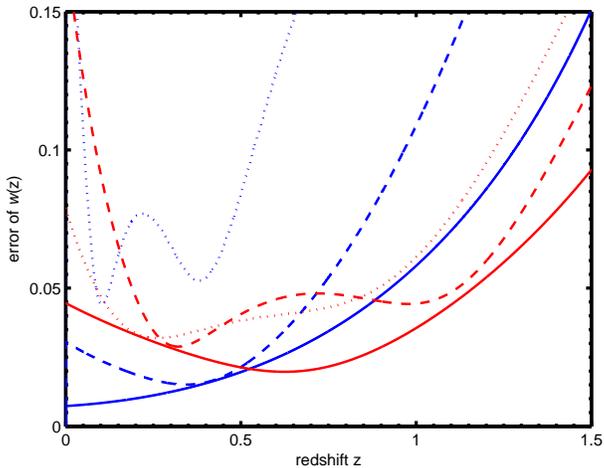}
\caption{Errors of $w(z)$ reconstructed from simulated data. Red curves correspond to
$H_s(z)$ data and blue curves correspond to $\mu(z)$ data. The solid, dashed, and dotted
lines denote $s$ = 1.0, 0.5, and 0.1 respectively.
\label{Fig:errsd}}
\end{figure}

For comparison a hypothetical data set of distance modulus $\mu(z)$
data is also constructed \cite{Clarkson2010104}, which consists of 2000 data
points evenly distributed in the redshift range $0.08 \leq z \leq 1.7$, and
300 data points in range $[0.03,0.08]$ \cite{Aldering2004523}. The uncertainties are
composed of a constant statistical error $\sigma_{\rm{mag}} = 0.15$
and systematic errors linearly drifting from $\sigma_{\rm{sys}} = 0
- 0.02$. The primary basis functions
$p_{n}(z) = [z/(1+z)]^{n-1}$, $n = 1\sim N$, are transformed again
to a set of eigenmodes with PCA, which are suitably adapted to the
simulated $\mu(z)$ data. The cosmological parameters $\{\Omega_{m},
\Omega_{k}, H_{0}\} = \{0.3, 0.0, 65.0\}$ are assumed, according to \cite{Clarkson2010104}. Different assumptions of $H_0$ in $H(z)$ and $\mu(z)$ simulations ($H_0$ is set to 67.4 and 65 respectively) don't affect the results.
Fig.\ref{Fig:errsd} demonstrates the
comparison between $w(z)$ errors reconstructed from $H_s(z)$
data and those reconstructed from simulated $\mu(z)$ data. Red
curves represent $H$-reconstructed errors and blue curves represent
$\mu$-reconstructed errors. We find that for moderate $z$ values
$(z\sim 0.4-0.8)$, both $H$-reconstructed errors and
$\mu$-reconstructed errors are relatively low. When $z$ is greater
than 0.6, however, the $H$-reconstructed errors are lower than
$\mu$-reconstructed ones, and with the increase of $z$, the
differences between them get more remarkable. Increasing $s$ also
vastly magnifies the differences. For smaller $s$, in contrast to the
$H$-reconstructed errors, the $\mu$-reconstructed errors will grow
uncontrollably as $z$ increases. Only in the low-$z$ region $(z < 0.4)$
does $H_s(z)$ data produce $w_{Hs}(z)$ errors greater than $\mu(z)$
data. In addition, both $\mu$-reconstructed and $H$-reconstructed
errors grow with decreased $z$ in the low-$z$ region $(z < 0.4$), and the
differences between them are acceptable. Taken together, Fig.\ref{Fig:errsd} (see also Fig.\ref{Fig:werrz1}) shows that we can constrain the $\Lambda$CDM model within $10\%$ at redshifts $z \lesssim 1.5$ and even $5\%$ at 0.1 $\lesssim$ z $\lesssim$ 1 by using the simulated $H(z)$ data.
Thus, we conclude that using $H_s(z)$ data can get better results than that using simulated
$\mu(z)$ data. Note that the mean relative errors of $H_s(z)$ data
are about $0.14\sim0.16$, while the relative errors of $\mu(z)$ data
are about $0.003\sim0.004$, indicating the former is nearly two
orders larger than the latter (we compare the relative errors rather
than the absolute values since $H_s(z)$ and $\mu(z)$ are different
physical quantities). Hence, the comparison convincingly validates
the merit of Hubble parameter data, as stated in section
\ref{merit}, in reconstructing the dark energy equation of state.

\begin{figure}
\centering
\includegraphics[width=9cm]{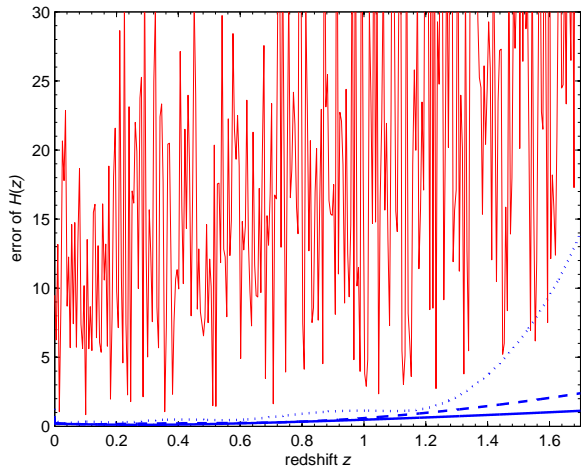}
\caption{Errors of $H_s(z)$ data (red) and errors of $H_r(z)$ data (blue). The solid, dashed, and dotted lines denote $s$ = 0.1, 0.5, and 1.0 respectively. Note that the errors produced by $H(z)$ simulation are random numbers.
\label{Fig:errh}}
\end{figure}

Based on the analysis in section \ref{merit}, the reconstruction of
$w(z)$ from $\mu(z)$ data can be split into two procedures:
first reconstructing $H(z)$ (denote by $H_r(z)$ hereafter) from $\mu(z)$ using Eq.(\ref{DtoH}), and then deriving $w(z)$
from $H_r(z)$ using Eq.(\ref{wzH}). One may imagine that the error propagation and
amplification taken in the first procedure will induce errors of
$H_r(z)$ similar in value to, or at least at the same order as,
errors of $H_s(z)$ data, so they would produce comparable errors
(see Fig.\ref{Fig:errsd}) for the
resulting $w(z)$. However, Fig.\ref{Fig:errh} shows this is not the case. We find from this figure that,
compared to the errors of $H_s(z)$ data, $H_r(z)$ errors are quite
small. We give the explanation as follows: first note that both
$H_s(z)$ and $H_r(z)$ data are noisy, with errors measuring the
extent of noise (see Fig.\ref{Fig:errh}). Because we use the first $M < N$ eigenfunctions,
which encompass the dominant features in the data, to fit the data,
and throw away the higher ones which contain noise-induced
oscillations (see Methodology), noises are suppressed in the resultant
$w_{Hs}(z)$. However, this is the case for $w_{Hs}(z)$
reconstruction from $H_s(z)$ data but not the case for
$w_{\mu}(z)$ reconstruction from $H_r(z)$ data. By resampling
the parameters in $\mu(z)$ data fitting, noises are to a large
extent suppressed in the first procedure of $w(z)$
reconstruction from $\mu(z)$ data (i.e., the $\mu(z)\rightarrow H_r(z)$ procedure), resulting in relatively low
$H_r(z)$ errors (see Fig.\ref{Fig:errh}), while the second procedure (i.e., the $H_r(z)\rightarrow w(z)$ procedure) essentially maintains all the noise-induced features inherited from the first procedure.
In a word, it is the noise-suppression effect in the reconstruction
of $w(z)$ from $H_s(z)$ data, which is lacking in the
second $H_r(z)\rightarrow w(z)$ procedure of $w(z)$ reconstruction from $\mu(z)$ data, that makes the
$w(z)$ reconstructed from $H_s(z)$ behave better, even though the
quality of $H_s(z)$ data is much worse than $H_r(z)$ data.

\begin{figure}
\centering
\includegraphics[width=4.0in]{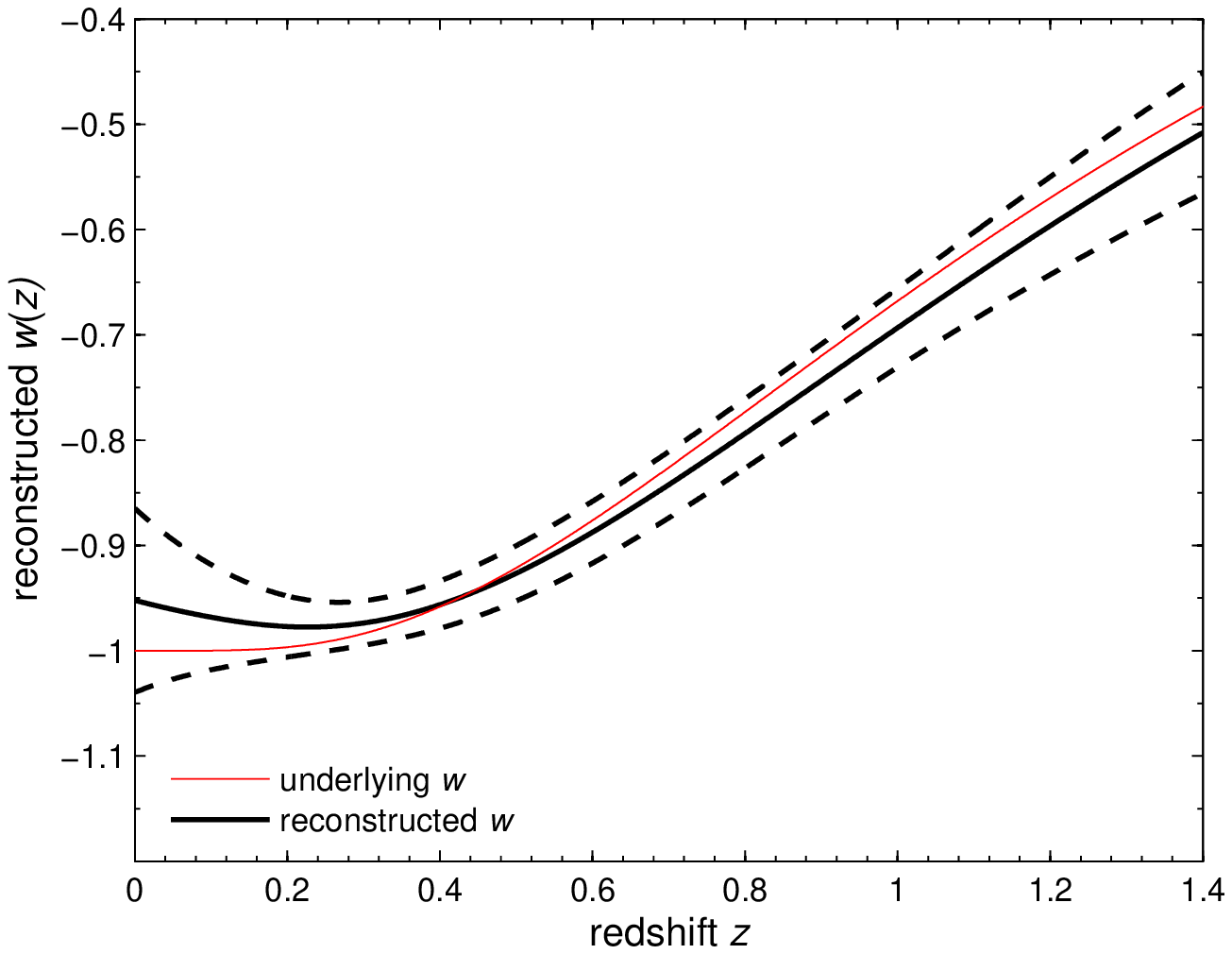}
\includegraphics[width=4.0in]{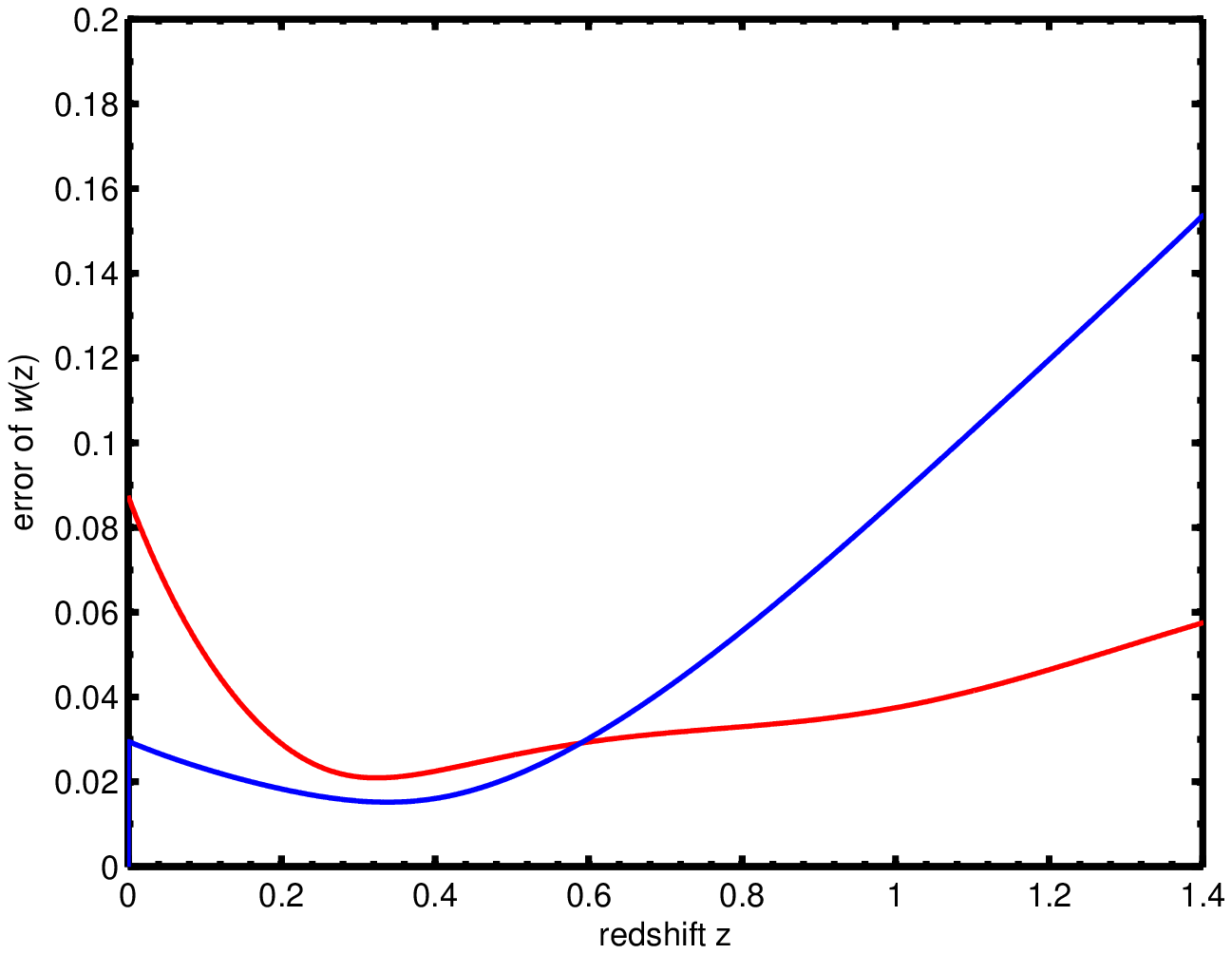}
\caption{Reconstruction of evolving dark energy of the first type using $H_s(z)$ data (a) and error comparison (b).
In (b), the red line denotes the errors constructed by $H_s(z)$ data and the blue line corresponds to simulated SN Ia data.
\label{Fig:hsimwf1}}
\end{figure}

\begin{figure}
\centering
\includegraphics[width=4.0in]{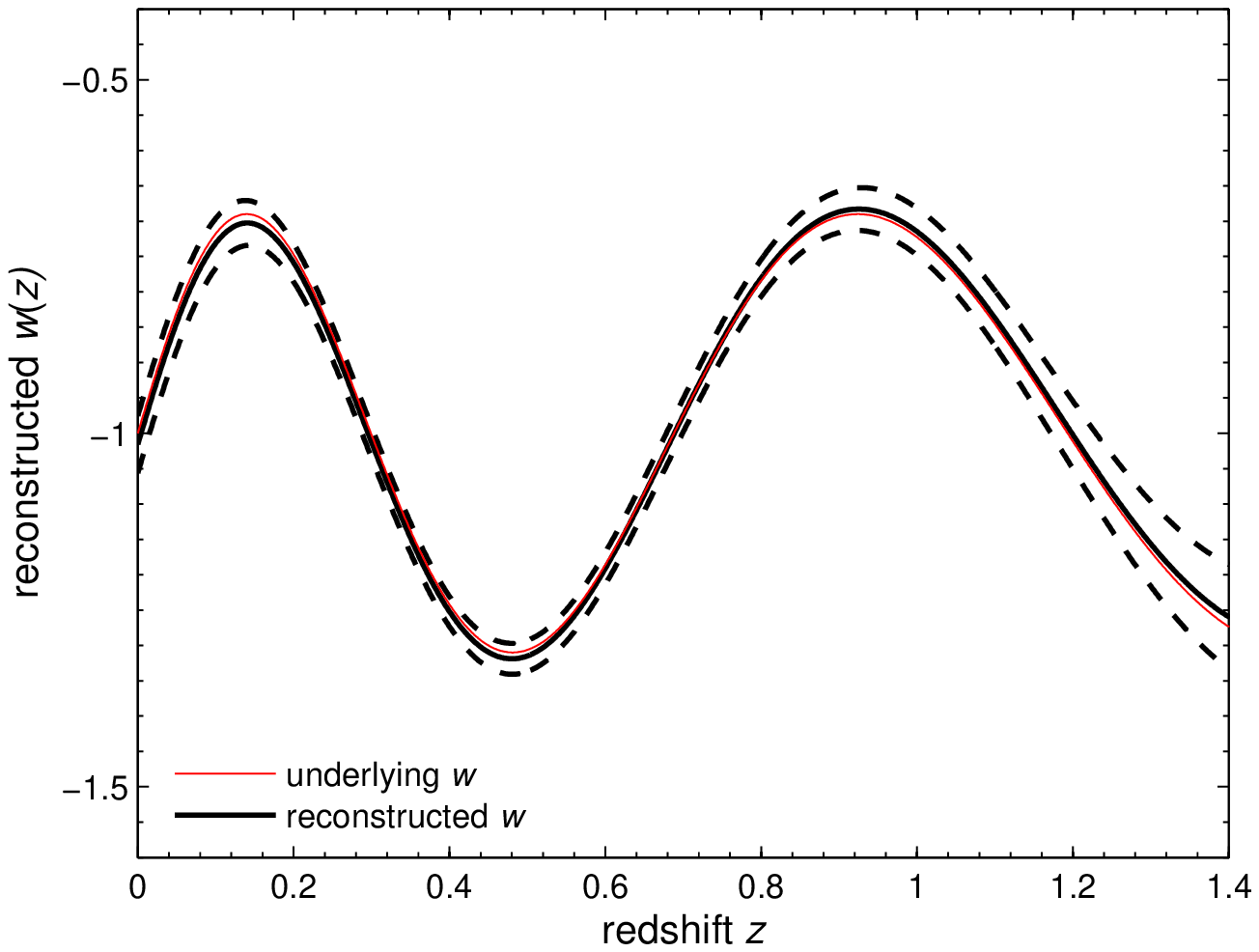}
\includegraphics[width=4.0in]{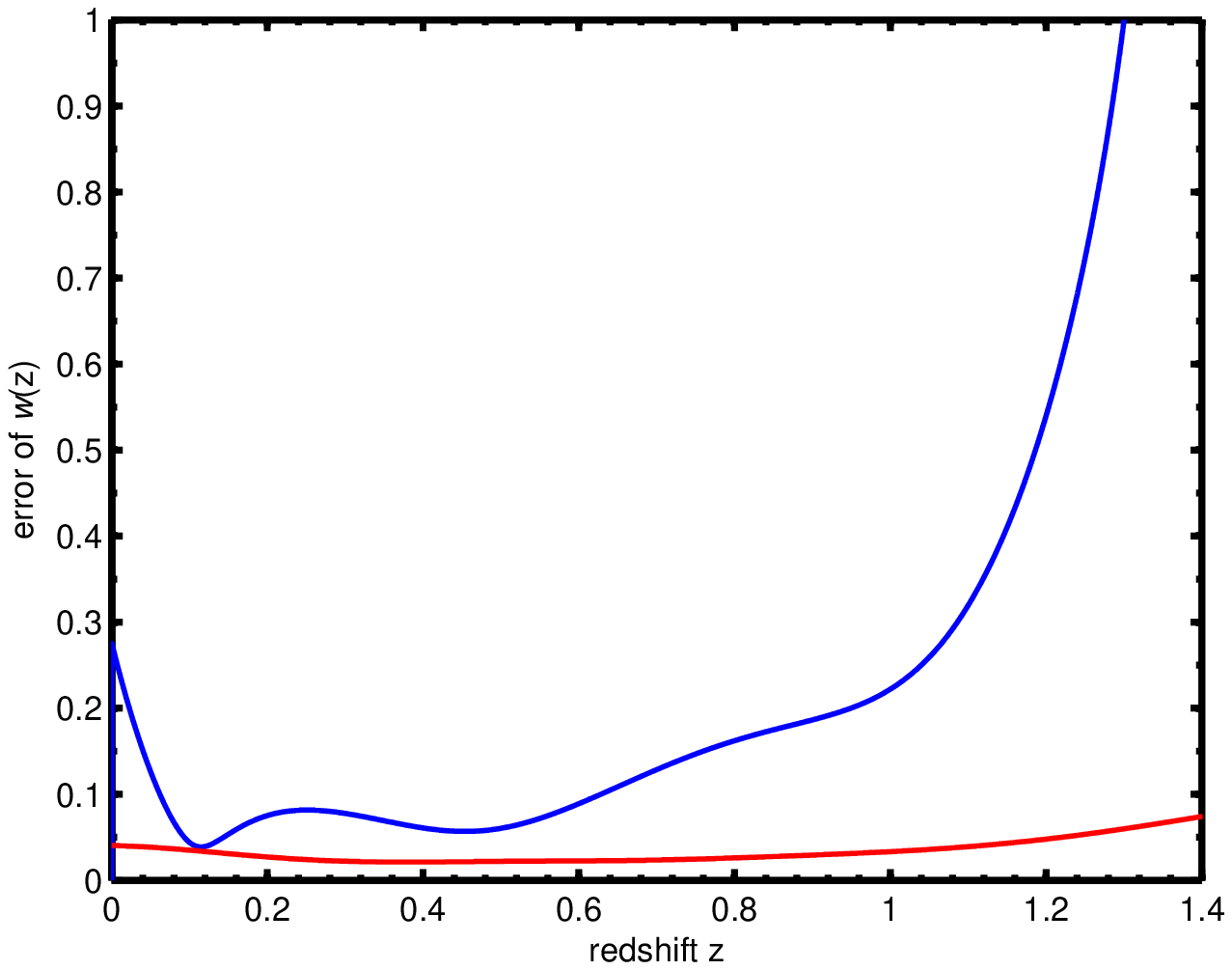}
\caption{Reconstruction of evolving dark energy of the second type using $H_s(z)$ data (a) and error comparison (b).
In (b), the red line denotes the errors constructed by $H_s(z)$ data and the blue line corresponds to simulated SN Ia data.
\label{Fig:hsimwf2}}
\end{figure}

Then we use $H_s(z)$ data to reconstruct two types of evolving $w_e(z)$. One is a standard
slow evolution, expressed as $w=\frac{1}{2}[-1+erf(\ln^{2z}-1)]$; the other evolves more drastically,
modeled by $w=-1+0.31\sin[12\ln^{(1+z)}]$. The reconstruction result is shown in Figs. \ref{Fig:hsimwf1}
and \ref{Fig:hsimwf2}. We set $s=0.2$ and $\{\Omega_{m}, \Omega_{k}, H_{0}\} = \{0.30, 0.0, 67.4\}$.
We again see that $H_s(z)$ data outperform, in most cases, the simulated SN Ia data. Similar to the result obtained from the $\Lambda$CDM model, the two evolving $w_e(z)$ models are also constrained within $10\%$ for most redshifts in the range $z \lesssim 1.5$.
For the second type of $w_e(z)$, not only the errors constructed by the former are much lower
than the latter, but also the former constructed $w_e(z)$ is more faithful to the underlying $w_e(z)$.
It seems that $H_s(z)$ data perform better for more oscillatory evolving dark energy.

 \subsection{Results on observational data}
 \label{resultobs}

\begin{figure}
\centering
\includegraphics[width=9cm]{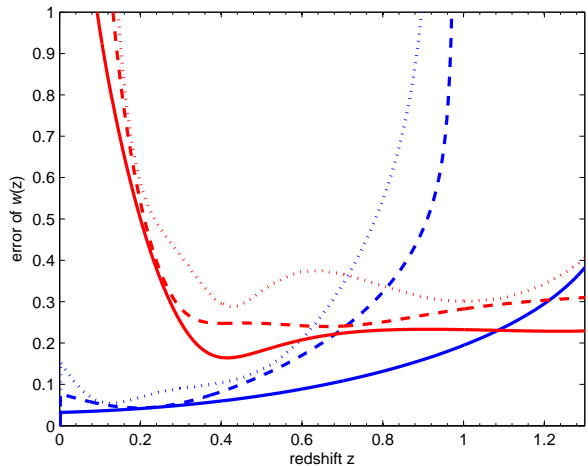}
\caption{Errors of $w(z)$ reconstructed from observational data. Red curves correspond to H data and blue curves correspond to $\mu$ data. The solid, dashed, and dotted lines denote $s$ = 1.0, 0.5, and 0.1 respectively.
\label{Fig:errod}}
\end{figure}

Finally we test the performance of OHD. See section 2 for the
description of the OHD dataset of 30 data points.
To make a comparison, the data set of distance modulus $\mu(z)$ is also
acquired. The last distance modulus measurements, which consist of
580 observed points, were produced by Supernova Cosmology
Project\footnote{\url{http://supernova.lbl.gov/Union/}}
\cite{Suzuki2012746}. The parameters of the best-fit fiducial $\Lambda$CDM model for OHD data are
$\{\Omega_{m}, \Omega_{k}, H_{0}\} = \{0.36, -0.13, 69.1\}$, while the parameters for fitting $\mu(z)$ data are $\{\Omega_{m}, \Omega_{k}, H_{0}\} = \{0.32, -0.09, 65.3\}$\cite{Clarkson2010104}. Interestingly, the most likely
$w(z)$ reconstruction, denoted
$w_{Ho}(z)$, favors a transition from $w_{Ho}(z)<-1$ at
low redshift to $w_{Ho}(z)>-1$ at higher redshift, a behavior
that is consistent with the quintom model which allows
$w_{Ho}(z)$ to cross -1 \cite{Feng2005607}.
The comparison of $1-\sigma$ errors
between $H$-reconstructed and $\mu$-reconstructed $w(z)$ is given in
Fig.\ref{Fig:errod}. We can see from Fig.\ref{Fig:errod} that the
quality of two reconstructions are comparable through most of the \textit{z}
region, though in low redshift region $H$-reconstructed errors are noticeably
larger than $\mu$-reconstructed errors. Roughly speaking, the
reconstruction from $H_o(z)$ performs better than the reconstruction
from $\mu(z)$ for relatively large $z$ values.
Concerning the much worse quality and much fewer observational data
points of $H_o(z)$ compared to $\mu(z)$, this result is impressive.

\begin{figure}
\centering
\includegraphics[width=3.0in]{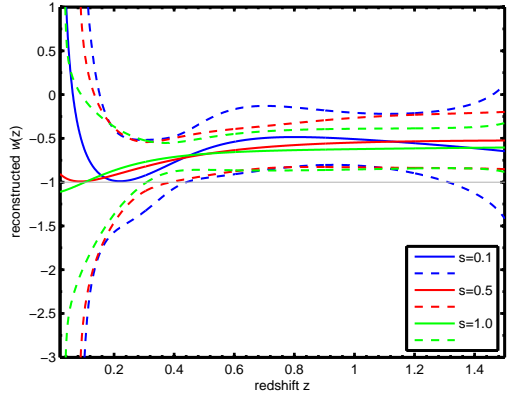}
\caption{ Reconstruction of $w$ using $H_o(z)$ data. The solid curves represent
reconstructed $w(z)$ and the dashed curves represent 1-$\sigma$ errors.
\label{Fig:ohd2w}}
\end{figure}

It is shown from Fig.\ref{Fig:ohd2w} that the shapes of the curves
$s=0.1$ and $s=0.5$ are quite similar. Accordingly, as shown in
Fig.\ref{Fig:errod}, the difference of error curves of
$w_{Ho}(z)$ for $s=0.1$ and $s=0.5$ are small, and the $s=1.0$
curve differs from the other two $H$-reconstructed curves at a lower
extent compared to the $\mu$-reconstruction counterparts, especially in high redshift region. This
behavior reveals that the reconstructions obtained from $H_o(z)$
data are more ``stable'' than those from $\mu(z)$ in the sense that
they are less dependent on $s$ values. Here we can give an intuitive
interpretation for it. The parameter $s$ in fact controls $M$ values
and the number of $\{N, M\}$ combinations. The smaller $s$ is, the higher the probability that the $Ms$ larger, and the bigger the
set of $\{N, M\}$ combinations is. Therefore, smaller $s$ results in
the emergence of larger $M$ in the set of $\{N, M\}$ combinations, which
leads to more precise data fitting and more noisy features reserved.
Since $\mu$ is more "distant" from $w$ than $H$, even small
uncertainties included in the SN Ia data fitting curves may induce
comparatively large oscillations of the resultant $w_{\mu}(z)$.
On the contrary, for the $w_(z)$ reconstruction from OHD, the noisy features entering the eigenmodes due to the
dropped \textit{s} will not cause the resultant $w(z)$ to
oscillate significantly. Hence, from this aspect the merit of OHD
in reconstructing dark energy equation of state is established more
solidly.

\section{Conclusions}
\label{conclusion}

We made an attempt in this paper to reconstruct dark energy equation of state parameter $w(z)$ from
Hubble parameter data within the framework proposed by Clarkson $\&$ Zuncke \cite{Clarkson2010104}. We first made a comparison between the simulated Hubble parameter $H_s(z)$ data and distance
module $\mu(z)$ data. The results showed that $H_s(z)$ data can produce more conservative $w(z)$
than the simulated $\mu(z)$ data. Then the performance of OHD was evaluated against SNe data. We found that the
errors of the reconstructed $w(z)$, using OHD and $\mu(z)$ data, are comparable. We also concluded that the
number of data points could impact the quality of the reconstruction. It turns out that 400 data points are sufficient to
improve the reconstruction quality using OHD only. Further improvement requires more accurate data measurements.

We have shown that the Hubble parameter dataset alone is potentially capable of reconstructing the dark energy equation of state. As discussed above, to obtain satisfactory $w(z)$ reconstruction, efforts on both the quality and the number of OHD are requisite. There are several ways to enhance the quality of OHD, of which deeper-redshift, more-complete-sky-coverage LRG survey and spectroscopic observations of those identified LRGs give remarkable improvement.
It is expected that future surveys will be able to offer a large sample of massive and passive galaxies to be used
in age-dating. The precise determination of the absolute galaxy ages is a quite complicated process, however, that issue can be ignored for $H(z)$ measurements because one just needs to take the difference between the ages in narrow redshift bins so that the systematic bias in the absolute ages cancel out. The ongoing CMB observation program, the Atacama Cosmology Telescope (ACT)~\cite{actweb}, will produce a catalogue of galaxy cluster candidates to be
validated up to $z\sim1.5$ via the Sunyaev-Zel'dovich effect \cite{Simon200571l, Lindner2015}. This ensures a promising dataset outperforming current SNIa datasets. For the range $1.5<z<2$, forthcoming surveys such as Euclid\cite{Laureijs2011} and WFIRST
\cite{Spergel2013}, will greatly improve the present-day statistics, increasing the number of passive galaxies at these redshifts by at
least an order of magnitude\cite{Laureijs2011}, which will result in observing approximatively one thousand spectra of passive galaxies at these redshifts\cite{Moresco2015}. Combining these future observations, it is reasonable to expect that the
Hubble parameter data will be helpful in exploring the expansion history of the Universe.

It should be noted that the results and analyses in this paper are method-specific, since we consider just one reconstruction method. Whether similar results and conclusions can be obtained by using other reconstruction methods, such as Gaussian process modeling \cite{Holsclaw2010105}, is still unknown.

\section*{Acknowledgements}

We would like to thank Dr. Hao-ran Yu and Hao-feng Qin for their in-depth discussion with us.
This work was supported by the National Science Foundation of
China (Grants Nos. 11173006, 11528306 and 11303003), the Ministry of Science and Technology National Basic Science
program (project 973, grant No. 2012CB821804) and the Shandong
Provincial Natural Science Foundation of China (Grant No ZR2013AM002).

\section*{References}

\bibliography{mybibfile}

\end{document}